\documentclass[amsmath,amssymb,aps,prb,reprint,superscriptaddress]{revtex4-1}
\usepackage{graphicx}
\usepackage{dcolumn}
\usepackage{bm}
\usepackage{multirow}

\begin{document}
\title{Edge states in lateral p-n~junctions in inverted band HgTe quantum wells}

\author{S.U.~Piatrusha}
\affiliation{Institute of Solid State Physics, Russian Academy of Sciences, 142432 Chernogolovka, Russian Federation}
\affiliation{Moscow Institute of Physics and Technology, Dolgoprudny, 141700 Russian Federation}
\author{V.S.~Khrapai}
\affiliation{Institute of Solid State Physics, Russian Academy of Sciences, 142432 Chernogolovka, Russian Federation}
\affiliation{Moscow Institute of Physics and Technology, Dolgoprudny, 141700 Russian Federation}
\author{Z.D.~Kvon}
\affiliation{Institute of Semiconductor Physics, Novosibirsk 630090, Russia}
\affiliation{Novosibirsk State University, Novosibirsk 630090, Russia}
\author{N.N.~Mikhailov}
\affiliation{Institute of Semiconductor Physics, Novosibirsk 630090, Russia}
\author{S.A.~Dvoretsky}
\affiliation{Institute of Semiconductor Physics, Novosibirsk 630090, Russia}
\author{E.S.~Tikhonov}
\affiliation{Institute of Solid State Physics, Russian Academy of Sciences, 142432 Chernogolovka, Russian Federation}
\affiliation{Moscow Institute of Physics and Technology, Dolgoprudny, 141700 Russian Federation}

\date{\today}

\begin{abstract}
We investigate lateral $p-n$~junctions, electrostatically defined in 14\,nm--wide HgTe-based quantum wells (QWs) with inverted band structure. The $p-n$~junctions resistances are close to $h/2e^2$, consistent with some previous experiments on $8$-$10\,\mathrm{nm}$ QWs, and the current-voltage characteristics are highly linear, indicating the transport via ballistic helical edge states. Shot noise measurements are performed in order to further verify the underlying transport mechanism.  We discuss the role of unknown inelastic relaxation rates in the leads and in the edge channels for the correct interpretation of the noise data. Although the interpretation in favor of the helical edge states seems more consistent, a definite conclusion can not be drawn based on the present experiment. Our approach looks promising for the study of short quasi-ballistic edges in topological insulators (TIs) in suitable geometry.
\end{abstract}

\maketitle

\section{Introduction}
Theoretical prediction for the possible existence of gapless edge states on the boundary between a topologically non-trivial and a conventional insulators~\cite{PhysRevLett.95.146802} has led to their experimental realization in several materials~\cite{Konig766,PhysRevLett.107.136603}. Concerning the HgTe QWs with inverted band structure, edge transport was verified, e.g., via non-local measurements~\cite{Roth294} or the visualisation of current density~\cite{Nowack2013,Hart2014}. Despite extensive study, however, some fundamental aspects are still lacking understanding. Edge conductance values, close to the quantum~$e^2/h$, were demonstrated only for the shortest edges no longer than few micrometers~\cite{Konig766,PhysRevLett.114.126802}. Up to now there is no agreement on the dominant scattering mechanism~\cite{Edgephysics2016,PhysRevLett.118.046801,PhysRevB.96.081405}, which leads to the linear length-dependence with the absent or weak dependence on temperature for the longer edges~\cite{PhysRevB.89.125305}. Moreover, recent experiments on InAs/GaSb QWs~\cite{PhysRevLett.117.077701,1367-2630-18-8-083005,PhysRevB.96.075406} demonstrate the possible coexistence of helical edges with trivial edge states, which are hard to be distinguished from each other by transport experiments. In the lack of clear demonstration of length-independent conductance of the edge equal to~$e^2/h$, further investigations are necessary to reveal the ballistic nature of the edge transport. Our objective here is to implement noise measurements for the study of edge states.
\par
The nature of edge conduction mechanism can be directly probed by the measurements of spontaneous current fluctuations (shot noise), which are induced by random scattering events. For example, for a strictly one-dimensional coherent conductor, the Fano-factor~$F\equiv S_I/2eI$, where $S_I$ is the spectral density of the fluctuations of the current~$I$, is directly related to the transmission~${\cal T}$ via $F=1-{\cal T}$. For a ballistic conductor the shot noise vanishes~\cite{PhysRevLett.75.3340,PhysRevLett.76.2778}, while in diffusive multi-channel regime $F=1/3$~\cite{NAGAEV1992103,PhysRevB.59.2871}. Despite the obvious potential of shot noise measurements, to the best of our knowledge there is only one experimental study considering current noise of the edge states~\cite{Tikhonov2015}. The authors examined relatively long resistive edges in $8\,\mathrm{nm}$ HgTe QWs and obtained~$F$ between~0.1 and~0.3, qualitatively consistent with the later developed theory~\cite{PhysRevB.94.045425}, which considers the exchange of electrons between edge states and conducting puddles in the insulating bulk of the sample. Up to now, vanishing shot noise -- the direct consequence and evidence of ballistic transport, was never observed for edges with $R_{\mathrm{edge}}\approx h/e^2$.
\par
Recently, the edge states contribution\cite{1367-2630-12-8-083058,0953-8984-26-8-085301} was experimentally identified in the conductance of lateral $p-n$~junctions in $8-10\,\mathrm{nm}$ HgTe QWs~\cite{Minkov2015}. Here, we investigate such junctions, defined electrostatically in 14\,nm HgTe QWs in the temperature range between~$60\,\mathrm{mK}$ and $0.8\,\mathrm{K}$. Transport measurements demonstrate that the bulk conduction is shunted by the edge channels on both sides of the $p-n$~junctions. The junction resistances close to $h/2e^2$, as well as their linear current-voltage characteristics, are the hall-mark signatures of the ballistic helical edge transport. We utilize the shot noise measurements in order to further distinguish between possible helical and trivial edge transport scenarios. For the noise approach we demonstrate the importance of (i)~minimizing the contacts resistance; (ii)~knowledge of the hole-phonon scattering rate~$\Sigma_{\mathrm{h-ph}}$ in p-type conduction region; (iii)~knowledge of the energy relaxation rate~$\Sigma_{0}$ in the edge. Comparing the experimental results with the model calculations we find more consistence in the case of helical edge states. Nevertheless, the uncertainty of various inelastic scattering rates involved does not allow to draw a definite conclusion. Our approach may allow one to infer the edge transport mechanism in samples designed more suitably for noise measurements or provided the inelastic rates are known in the future.

\section{Samples and measurement technique} 
Our samples are based on 14\,nm wide (112)~CdHgTe/HgTe/CdHgTe QWs grown by molecular beam epitaxy, with mesas shaped by wet etching and covered with a SiO$_2$/Si$_3$N$_4$ insulating layer, $200\,\mathrm{nm}$ in total, see Ref.~\cite{Kvon2011} for the details. Metallic Au/Ti top gates enable us to tune the 2D~system across the charge neutrality point~(CNP) by means of field effect. Ohmic contacts are achieved by a few second In soldering in air, providing a typical resistance of the ungated mesa arms $\approx5\,\mathrm{k\Omega}$ at low~$T$. The experiment was performed in a dry Bluefors dilution refrigerator down to electronic temperature of~$60\,\mathrm{mK}$ (verified by noise thermometry) and in a liquid~$^3$He insert with a bath temperature $T=0.5\,\mathrm{K}$. Transport measurements were performed in a two-terminal or multi-terminal configurations with a low-noise $100\,\mathrm{M\Omega}$ input resistance preamplifier. The voltage fluctuations were measured within a frequency band $20-24\,\mathrm{MHz}$ for sample~S1 and within $8.5-9.5\,\mathrm{MHz}$ for sample~S2 on a $R_0=10\,\mathrm{k\Omega}$ load resistor, connected in parallel with $22\,\mathrm{MHz}$ ($9\,\mathrm{MHz}$) resonant tank circuit. The signal was amplified by a home-made $10\,\mathrm{dB}$ low-temperature amplifier (LTAmp), followed by a $3\times25\,\mathrm{dB}$ room-$T$ amplification stage, and measured with a power detector (not shown in fig.~\ref{fig1}a). The setup was calibrated via the Johnson-Nyquist thermometry. Below we present the results obtained on two samples which demonstrate similar behavior reproducible with respect to thermal recycling. We note that transport measurements were also performed on two more samples with similar results.

\section{Transport measurements}
\begin{figure}[t]
\begin{center}
\vspace{0mm}
\includegraphics[width=0.8\columnwidth]{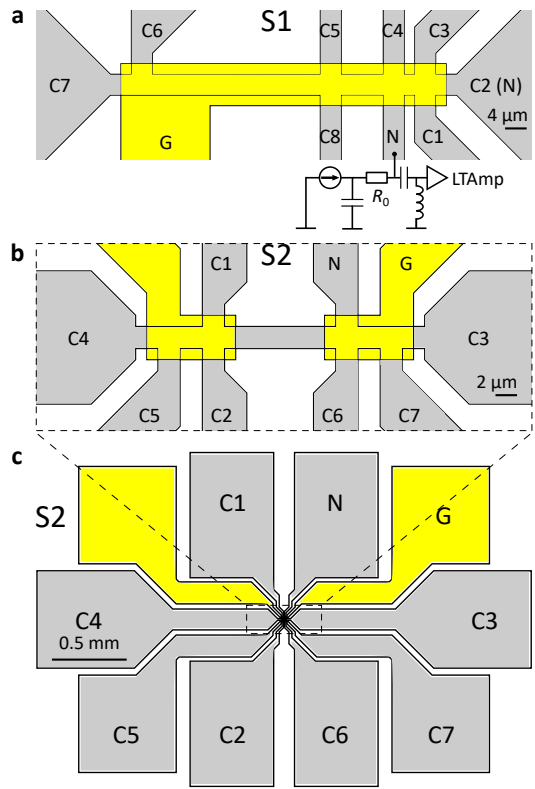}
\end{center}
\caption{Sample schematics. (a,b)~Schematic representation of the central part of the samples~S1 and~S2 with measurement scheme. The noise was measured from contacts N on both samples and additionally from contact~C2 on sample~S1. (c)~The layout of the whole sample~S2.}\label{fig1}
\end{figure}

The samples layouts are shown schematically in fig.~\ref{fig1} with 2D~leads coloured in gray and electrostatic gates in yellow. Sample~S1 is equipped with one large gate, while sample~S2 is equipped with two smaller gates, only one of which (labelled with~G) was used in the present experiment. Via indexed letters~\textquoteleft Ci\textquoteright\, we denote 2D~mesa leads (see the sketch of the whole sample~S2 in fig.~\ref{fig1}c). Contacts labelled with N were used for noise measurements. From four-point resistance measurements at gate voltages~$V_{g}=0\,\mathrm{V}$ and~$V_{g}=-6\,\mathrm{V}$ we extract the electron (hole) resistivities~$\rho_e$ ($\rho_h$) of $230\,(2000)\,\mathrm{\Omega}$ for sample~S1 and $240\,(3200)\,\mathrm{\Omega}$ for sample~S2. 
\begin{figure}[t]
\begin{center}
\vspace{0mm}
\includegraphics[width=0.9\columnwidth]{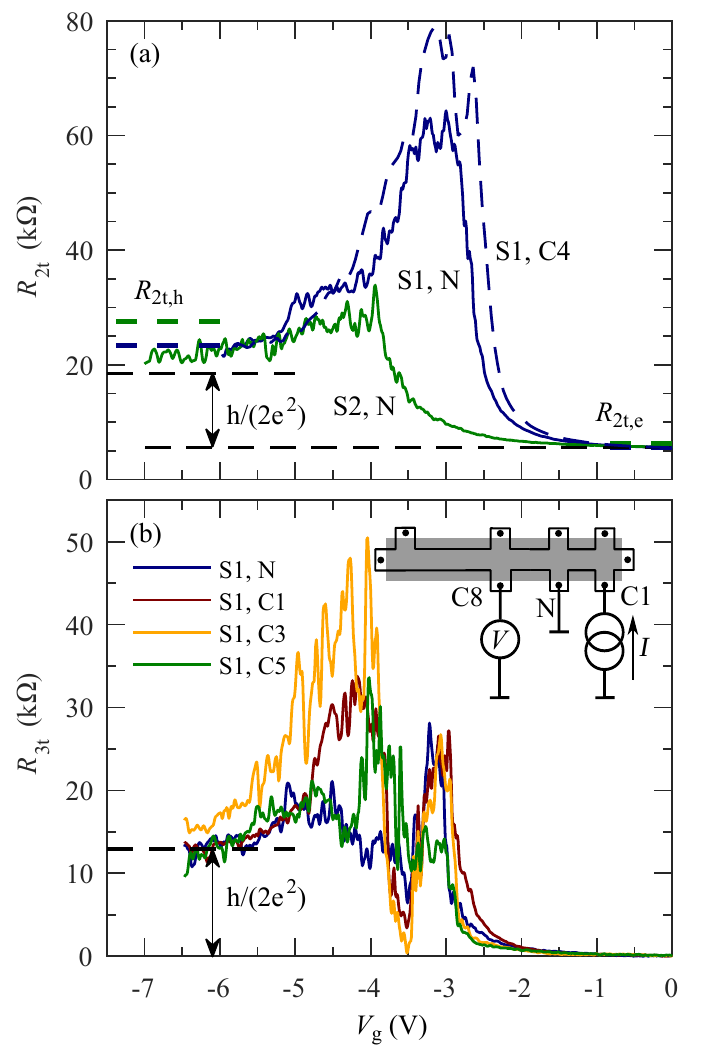}
\end{center}
\caption{Local transport measurements. (a)~Two-terminal resistance at~$T=0.5\,\mathrm{K}$ as a function of gate voltage for indicated contacts. The dashed curve (contact S1, C4) is shifted downwards by $2.8\,\mathrm{k\Omega}$. (b)~Three-terminal resistance as a function of gate voltage for different contacts obtained on the sample S1 at~$T=80\,\mathrm{mK}$. All curves are vertically offset by the corresponding value at~$V_{g} = 0\,\mathrm{V}$. Inset: layout of a three-terminal measurement in sample S1 for the $p-n$~junction on the N-lead.}\label{fig2}
\end{figure}
\par
In fig.~\ref{fig2}a we plot the gate voltage dependencies of the two terminal linear response resistances ($R_{2t}$) for several contacts on both samples~S1 and~S2. In these measurements all contacts, except for the studied one, are grounded. At~$V_{g}=0\,\mathrm{V}$ the measured value, $R_{2t,e}$, reflects the resistance of 2D~mesa leads and ohmic contacts at the In/2DEG interface (which we denote via $R_{\mathrm{Ci/N,ohmic}}$). This value is on the order of a few kiloohms for most of the contacts. However, contacts~C4 and~C5 of sample~S2 were not not working, probably, due to poor In soldering, and contacts~C6 and~C7 of sample~S2, as well as contact~C3 of sample~S1, had relatively large resistances of approximately~$20\,\mathrm{k\Omega}$, $70\,\mathrm{k\Omega}$ and~$20\,\mathrm{k\Omega}$, respectively.
\par
At negative enough~$V_{g}$, when p-type conduction regime is set up in the mesa region under the gate, the measured resistance, $R_{2t,h}$, is determined simultaneously by ohmic contacts, 2D~leads, p-type conduction area under the gate and $p-n$~junctions, formed in the mesa under the gate edge. For all samples studied, $R_{2t,h}$ is almost $V_{g}$-independent, though with visible fluctuations, at~$V_{g}<-5\,\mathrm{V}$. While both~$R_{2t,e}$ and~$R_{2t,h}$ values may be affected by the unknown ohmic contact resistance, their difference eliminates this uncertainty and provides information about $p-n$~junctions resistances. We note that for most contacts for all samples the difference $R_{2t,h}-R_{2t,e}$ is always close to~$R_Q\equiv h/2e^2$ (see dotted lines in fig.~\ref{fig2}), exceeding it by several kiloohms, which may correspond to the resistance of under-the-gate p-type conduction region. This observation indicates that the transport in $p-n$~junctions occurs predominantly via two ballistic helical edge channels on both sides.
\par
Assuming all $p-n$~junctions have resistance~$R_{\mathrm{pn}}=R_Q$ and using measured values of~$\rho_{e,h}$, for geometry corresponding to mesa mask we calculate the expected two terminal resistance (for both~n- and p-type conduction under the gate) for different contacts, and compare it with experimental values. The expected values are illustrated in fig.~\ref{fig2}a by thick dashed straight lines. For contact~N on sample~S1 the result is perfectly consistent with the measured values both at~$V_{g}=0\,\mathrm{V}$ and~$V_{g}=-6\,\mathrm{V}$ under the assumption of vanishing corresponding ohmic contact resistance~$R_{\mathrm{N,ohmic}}=0\,\Omega$. For contact~C4 on sample~S1 both measured resistances are greater than calculated values by the same amount of $\approx2.8\,\mathrm{k\Omega}$, which most likely reflects~$R_{\mathrm{C4,ohmic}}=2.8\,\mathrm{k\Omega}$. This consistency shows that the resistance of $p-n$~junctions in our devices is indeed close to resistance quantum. For contact~N on sample~S2, while the agreement for n-type conduction is good with negligibly small~$R_{\mathrm{N,ohmic}}=0\,\Omega$, the measured resistance for p-type conduction under the gate is by $\sim4\,\mathrm{k\Omega}$ smaller than the calculated one. This might indicate the bulk of $p-n$~junction contribution to transport.
\par
To explicitly demonstrate that the values of $R_{\mathrm{pn}}$ are close to the quantized value, we perform the three-terminal resistance measurements. As an example, the corresponding measurement scheme for contact~N is illustrated schematically in the inset of fig.~\ref{fig2}b. This geometry allows to minimize the poorly known input of area under the gate. In fig.~\ref{fig2}b we plot the change $R_{3t}(V_{g})-R_{3t}(V_{g}=0)$ of the measured value as a function of~$V_{g}$. The data demonstrate that indeed at negative enough gate voltages this change is close to~$R_Q$, up to the fluctuations. We note that the observation $R_{\mathrm{pn}}\approx R_Q$ holds also for samples where noise measurements were not performed and persists in all studied temperature range.
\par
\begin{figure}[t]
\begin{center}
\vspace{0mm}
\includegraphics[width=0.9\columnwidth]{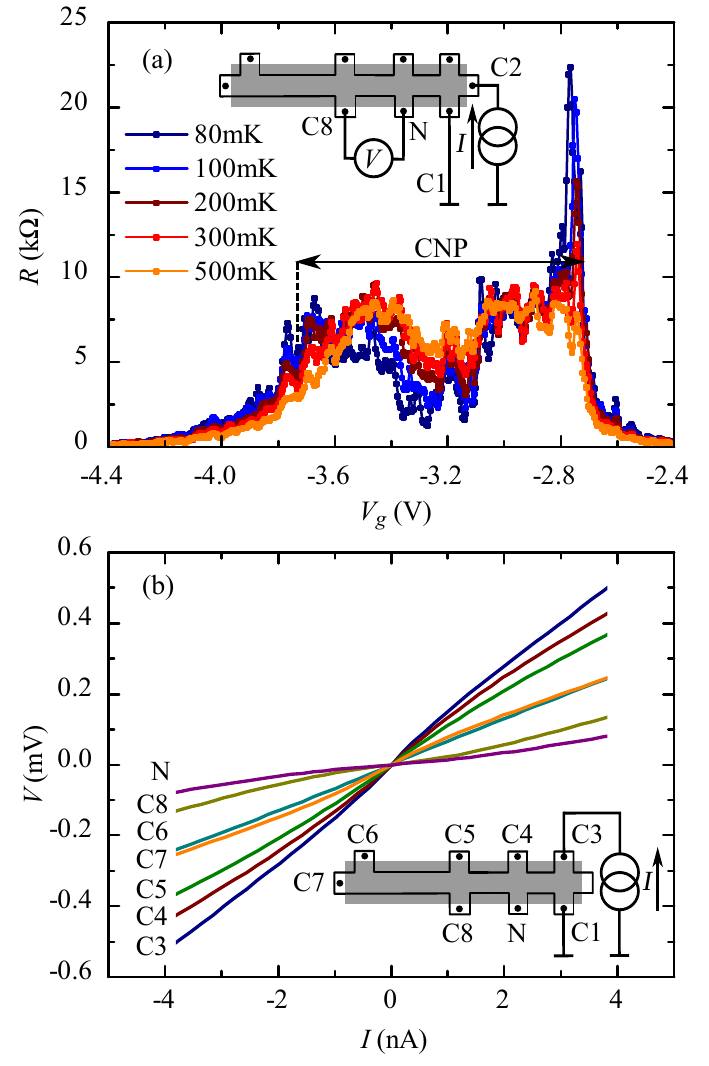}
\end{center}
\caption{Evidence of the edge transport in the linear and non-linear response regime. (a)~Non-local resistance as a function of gate voltage at different temperatures. (b)~Edge transport regime for~$V_{g} = -3.1\,\mathrm{V}$ in the CNP region at~$T=80\,\mathrm{mK}$.}\label{fig3}
\end{figure}

Edge transport in $14\,\mathrm{nm}$-wide QWs in similar samples was studied recently~\cite{PhysRevLett.114.126802}. The presence of edge states in our experiments in both linear and non-linear transport regimes is verified in fig.~\ref{fig3}. In fig.~\ref{fig3}a we plot the non-local linear-response resistance, measured as indicated in the inset, as a function of~$V_{g}$ at different temperatures in the range $80\,\mathrm{mK}-0.5\,\mathrm{K}$. Qualitatively, the behavior of local (fig.~\ref{fig2}a) and non-local resistances is similar and demonstrates the gate voltage range in which the transport occurs via edge channels. Fig.~\ref{fig3}b shows the set of three terminal $I$-$V$~curves measured on sample~S1 as indicated in the inset. Here, the source contact is C3, the ground contact is C1 and the voltage on the contact probes~C4--C8 and N is measured in respect to the ground potential (see Fig.~\ref{fig1} for contacts labels; contact~C2 was not soldered in this experiment). These non-local $I-V$~curves measured with different contacts vary considerably: the dependence of~$V$ on the contact position corresponds to the contacts counterclockwise order, reflecting that near the CNP the transport current in our samples flows around the mesa edges under the gate. Moving away from the source contact, we observe the transition from sub-linear (which is typical for the measurements of local resistance, see Appendix~\ref{DR}) to super-linear $I$-$V$~behavior, probably indicating the bulk contribution to transport at distances bigger than~$\approx40\,\mathrm{\mu m}$ in the considered bias voltage range.
\par
For relatively long edges in the present samples we estimate the edge resistivity $\rho_{\mathrm{edge}}\lesssim25\,\mathrm{k\Omega/\mu m}$. The observation $R_{\mathrm{pn}}\approx R_Q$ then implies that the $p-n$~junctions length $L_{\mathrm{pn}}\gtrsim 1\,\mathrm{\mu m}$, which is much longer than the expected value of the length of the depleted bulk region in the $p-n$ junction~$\sim200\,\mathrm{nm}$. This observation further suggests that the edge resistance is independent of the length below $1\,\mathrm{\mu m}$, and hence the realization of transport across the $p-n$~junctions via one-dimensional helical edge states. The observation $R_{\mathrm{pn}}\approx R_Q$ then suggests the suppression of backscattering (e.g. due to charge puddles in the bulk of the $p-n$~junctions~\cite{PhysRevB.88.165309,PhysRevX.3.021003,PhysRevB.94.045425}	) on the length of the $p-n$~junction. In the following we utilize the shot noise measurements in order to distinguish between possible helical and trivial edge transport scenarios. We will compare the obtained results with two limit predictions, corresponding to either ballistic or diffusive conduction across the $p-n$~junction. 

\section{Noise measurements}
\begin{figure}[h]
\begin{center}
\vspace{0mm}
\includegraphics[width=0.9\columnwidth]{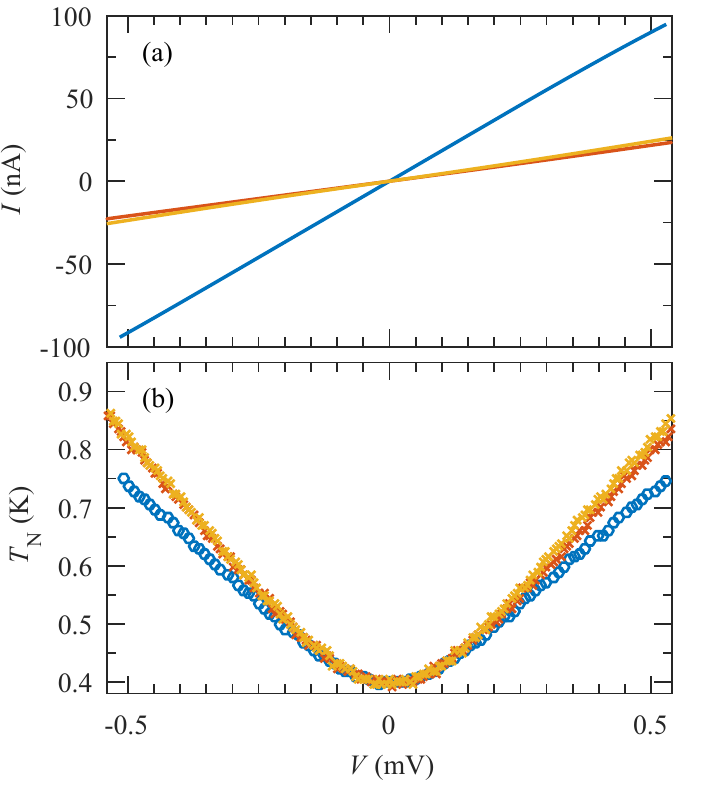}
\end{center}
\caption{I-V and noise measurements on the sample~S1 at $T=0.4\,\mathrm{K}$. (a)~I-V curves measured in a two-terminal configuration from the contact~$N$ at $V_{g}=0\,\mathrm{V}$ (blue) and at large negative $V_{g}=-6.175\,\mathrm{V}$ and $V_{g}=-6.5\,\mathrm{V}$. (b)~Noise temperature at $V_{g}=0\,\mathrm{V}$ (circles) and at $V_{g}=-6.175\,\mathrm{V}$ and $V_{g}=-6.5\,\mathrm{V}$ (crosses).}\label{fig4}
\end{figure}

In fig.~\ref{fig4} we plot the $I-V$~curves and noise temperature~$T_N=S_IR/4k_B$ obtained at $V_{g}=0V$ and large negative values $V_{g}=-6.175\,\mathrm{V}$ and $V_{g}=-6.5\,\mathrm{V}$, with $p-n$~junctions formed, from contact~N of sample~S1 at $T=0.4\,\mathrm{K}$. We note that the two-terminal differential resistance of our samples at large negative gate voltages is almost bias-independent, with changes on the order of few percent, and is symmetric in respect to bias inversion (see fig.~\ref{fig4}a). This behavior is in contrast with textbook behavior of semiconductor $p-n$~junctions and further indicates the edge states shunting the bulk of the $p-n$~junctions. At large negative~$V_{g}$, noise temperature~$T_N$ is almost $V_{g}$-independent and is higher than at~$V_{g}=0\,\mathrm{V}$, viewed as a function of total applied bias voltage (see fig.~\ref{fig4}b). We emphasize, that in the presence of relatively resistive contact leads this is not necessarily due to the shot noise contribution of the $p-n$~junctions and may simply reflect the well-known overheating effect, as discussed below.
\par
In the absence of electron-phonon (e-ph) relaxation, the shot noise spectral density is conveniently expressed in units of the Schottky value $F\equiv S_I/2eI$. Universal theoretical values of the Fano-factor~$F=0$ and~$F=1/3$ (or~$F=\sqrt3/4$ in the case of strong electron-electron (e-e) scattering~\cite{PhysRevB.52.4740,PhysRevB.52.7853,PhysRevLett.76.3806}) obtained for ballistic and diffusive conductors, respectively, hold assuming ideal boundary conditions for the electrons immediately at the sample ends, i.e. for reservoirs of infinite size with infinite electric and heat conductivities~\cite{PhysRevB.59.2871}. Experimentally, however, the measured signal may be significantly affected by overheating of the leads, which depends on its material, actual size and heat conductance. Qualitatively, in the presence of current flow, the Johnson-Nyquist noise is no longer constant with current but grows with bias. For example, for the otherwise noiseless ballistic conductor connected in series with realistic supply leads, this overheating is twofold. First, the temperature of the contact 2D~leads itselves grows in accordance with the total dissipated Joule heat. This, in turn, modifies the electronic distribution function at the ends of the ballistic conductor and thus leads to further noise increase. This effect was observed as the measured noise increase for diffusive wires connected to reservoirs with insufficient heat conductance~\cite{PhysRevB.59.2871}, or as finite noise on the conductance plateaus of ballistic quantum point contacts~\cite{PhysRevLett.75.3340,PhysRevLett.76.2778,PhysRevB.90.161405,PhysRevB.93.195411}. The magnitude of this stray noise temperature increase is determined by the ratio of conductor's and 2D~leads resistances. Concerning available shot noise experiments on HgTe QWs, while overheating effect is not significant for long resistive edges ($R_{\mathrm{edge}}/R_{\mathrm{cont}}\approx10$)~\cite{Tikhonov2015}, it is not the case in the present study ($R_{\mathrm{pn}}/R_{\mathrm{cont}}\lesssim3$). As a result, the proper consideration of reservoir noise is of great importance here. In the following, we first study reservoirs heating in response to the flowing current and then discuss the obtained results for the noise of $p-n$~junctions.

\subsection{Contacts heating}
\begin{figure}[h]
\begin{center}
\vspace{0mm}
\includegraphics[width=.9\columnwidth]{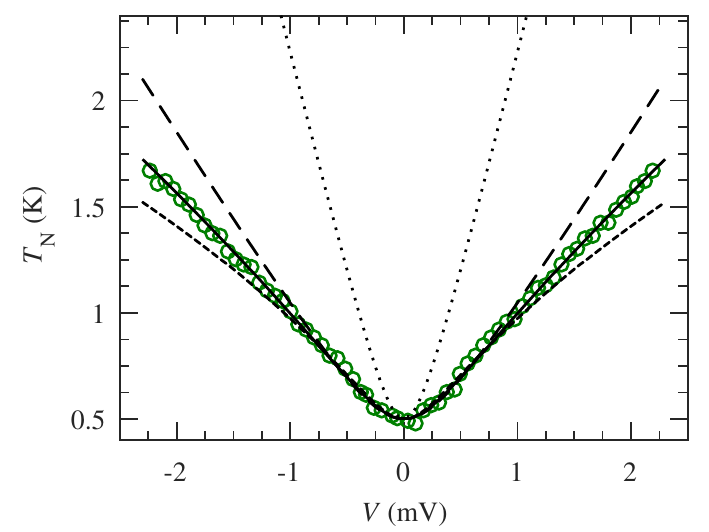}
\end{center}
\caption{Contacts heating study on the sample~S2. Noise temperature as a function of bias voltage (symbols) at~$V_g=0\,\mathrm{V}$. Dotted line is the prediction for the case when e-ph scattering is absent. Dashed, solid and short-dashed lines correspond to power-law heat transfer between electron and phonon systems with exponent~$\alpha=2$, 3 and 4, respectively.}\label{fig5}
\end{figure}

In order to take the influence of the n-type contacts into account, we set~$V_g=0\,\mathrm{V}$ and study the effect of 2D~leads heating in response to the flowing current. Experimental results are shown in fig.~\ref{fig5} for sample~S2 by circles. To demonstrate the influence of inelastic e-ph scattering, by dotted line we show the prediction for the case, when e-ph scattering is absent. The obvious suppression of the measured~$T_N$ reflects the fact that thermal relaxation of electrons is realized not only by electronic diffusion towards the ohmic contacts, but also by e-ph coupling. Local power flow between electron and phonon systems is generally given by
\begin{equation*}
P_{\mathrm{e-ph}}=\Sigma_{\mathrm{e-ph}} (T^{\alpha}-T_0^{\alpha}),
\end{equation*}   
where $T=T(x,y)$ -- position-dependent electronic temperature, $T_0$ -- bath temperature and $\Sigma_{\mathrm{e-ph}}$ is material-dependent e-ph coupling coefficient. Exponent~$\alpha$ characterizes the heat transfer mechanism and in general varies in the range~$\alpha\approx3-5$~\cite{RevModPhys.78.217}.
To evaluate~$\Sigma_{\mathrm{e-ph}}$ and~$\alpha$ in our case, we first numerically solve the two-dimensional stationary heat diffusion equation
\begin{equation*}
-\nabla (\varkappa \nabla T)=Q_{\mathrm{Joule}}-P_{\mathrm{e-ph}}
\end{equation*}
along with Poisson's equation for electric potential, for the sample shape corresponding to the lithographic mesa pattern. Here $Q_{\mathrm{Joule}}=\mathbf{j\cdot E}$ -- local Joule heat power source, ${\cal L}=2.44\times10^{-8}\,\mathrm{W\Omega K^{-2}}$ -- the Lorenz number, and electronic heat conduction is assumed to satisfy the Wiedemann-Franz law
\begin{equation*}
\varkappa=\sigma {\cal L} T.
\end{equation*}
Using thus obtained electronic temperature profile, we calculate the noise temperature of the sample via (see Appendix~\ref{BP})
\begin{equation}
T_N=\frac{\int T\, Q_{\mathrm{Joule}}\,\mathrm{d}x\mathrm{d}y}{\int Q_{\mathrm{Joule}}\,\mathrm{d}x\mathrm{d}y},
\label{BPF}
\end{equation}
with integration taken over the whole mesa. 
\par
We find that in the bias voltage range of few millivolts experimental data are best fitted with~$\alpha\approx3$ (see fig.~\ref{fig5}, solid line). The corresponding value of~$\Sigma_{\mathrm{e-ph}}=0.016\,\mathrm{W/m^2K^3}$ is the same for both samples~S1 and~S2 and is reproducible with respect to thermal recycling. Passing, we note that for larger bias voltages up to~$\sim100\,\mathrm{mV}$ (corresponding to $T_N\lesssim15\,\mathrm{K}$) the data are perfectly consistent with~$\alpha=4$ and~$\Sigma_{\mathrm{e-ph}}\approx0.007\,\mathrm{W/m^2K^4}$. Without going into the details of the underlying microscopic mechanism leading to $\alpha=3$, we note that it was observed, e.g. in graphene~\cite{PhysRevLett.109.056805,Betz2012}. Concerning our further analysis, we emphasize that the specific exponent is not important as long as it adequately describes the heating of contacts in the required range of~$T_N$.

\subsection{Shot noise of $p-n$~junctions}
\subsubsection{Model for simulations}
\begin{figure}[h]
\begin{center}
\vspace{0mm}
\includegraphics[width=0.8\columnwidth]{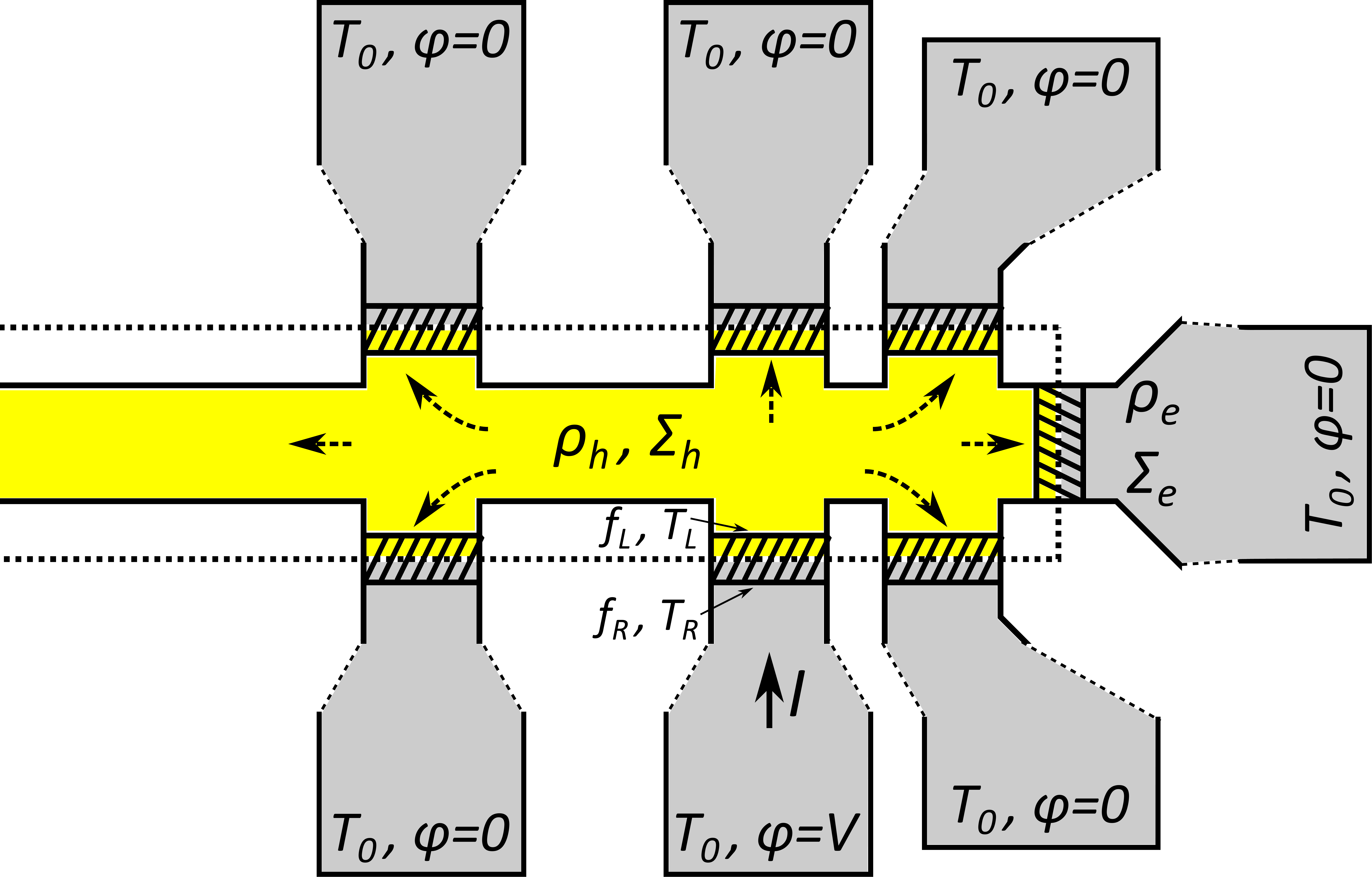}
\end{center}
\caption{Model for calculations. P-type conduction region is yellow, n-type conduction region is gray.  Dotted line shows the gate boundary. We fix the temperatures~$T=T_0$ and electric potentials~$\varphi_N=V$, $\varphi_{Ci}=0$ at ohmic contacts.}\label{fig6}
\end{figure}

In the following we will compare experimental results for the noise of $p-n$~junctions with numerical simulations obtained in the model which we now describe. Our model geometry (see fig.~\ref{fig6}) corresponds to mesa mask geometry with dotted rectangular indicating the gate position. Gray areas are the n-type conduction 2D~leads with electron resistivity~$\rho_e$ and e-ph coupling~$\Sigma_{\mathrm{e-ph}}$, evaluated above; yellow area is the p-type conduction region under the gate with hole resistivity~$\rho_h$ and unknown hole-phonon coupling~$\Sigma_{\mathrm{h-ph}}$. Shaded rectangles at the edge of the gate model $p-n$~junctions, each with conductance~$G=G_Q\equiv2e^2/h$, which are formed at large negative gate voltages. We set boundary conditions by defining temperatures~$T=T_0$ and electric potentials~$\varphi_N=V$, $\varphi_{Ci}=0$ at the ohmic contacts. The bias voltage, $V$, applied to the N-contact, defines the current through each mesa arm, allowing the temperature map to be obtained from the combined solution of Poisson's and heat diffusion equations. This is enough to calculate the noise temperature of the sample as we discuss below. 
\par
Generally, noise power of the current fluctuations in the non-interacting scattering theory is determined by two inputs~\cite{Blanter2001}:
\begin{equation*}
S=S_{\mathrm{thermal}}+S_{\mathrm{partition}}.
\end{equation*} 
Here, the first term is the analog of the equilibrium noise contribution (i.e. the Johnson-Nyquist noise), present in any conductor --
\begin{equation*}
S_{\mathrm{thermal}}=G_Q\sum\limits_{n}\int\mathrm{dE}\left\{T_n\left[f_L(1-f_L)+f_R(1-f_R)\right]\right\}.
\end{equation*}
Note that the electronic energy distributions~$f_L$ and $f_R$ on the left and the right hand sides of the conductor are not necessarily the same. The second term is the shot noise contribution that originates from partitioning of the charge current in the conductor --
\begin{equation*}
S_{\mathrm{partition}}=G_Q\sum\limits_{n}\int\mathrm{dE}\left\{T_n(1-T_n)\left(f_L-f_R\right)^2\right\}.
\end{equation*}
\par
We consider two limit predictions for the noise of our $p-n$~junctions. The first one corresponds to ballistic conduction across the $p-n$~junction, when for the $p-n$~junctions $S_{\mathrm{partition}}$ vanishes and their current spectral density is completely determined by~$S_{\mathrm{thermal}}$:
\begin{equation*}
S_{\mathrm{ball}}=2G_Qk_B(T_L+T_R).
\end{equation*}
Here the factor $2$ comes from the fact that conduction is realized along two edges of the junction. Additionally, we assume electronic energy distributions at the boundaries of the $p-n$~junction to be Fermi-Dirac with temperatures~$T_L$ and $T_R$, that are obtained numerically from the solution of heat diffusion equation. Such an assumption ensures the proper heat flow outwards the $p-n$~junction in our model calculations. 
\par
Another limit is realized when conduction across the $p-n$~junction is diffusive (in the case of trivial edge states). Here, current spectral density is determined by both equilibrium and shot noise contributions. In principle, two situations are possible for diffusive conduction. (i)~In the case of strong enough e-e scattering, when each point inside the junction may be considered as in local equilibrium with position-dependent temperature, eq.(\ref{BPF}) holds and current noise of each $p-n$~junction can be calculated via
\begin{equation}
S_{\mathrm{diff}}=4k_BT_{N,\mathrm{pn}}G_Q.
\label{localeq}
\end{equation}
With negligible e-ph~scattering this relation is equivalent to the familiar result $T_{N,\mathrm{pn}}=FeV_{\mathrm{pn}}/2k_B$ with $F=\sqrt3/4$~\cite{PhysRevB.52.4740}. (ii)~If the $p-n$~junction is short enough so that e-e scattering is negligible, another situation might be the case -- electron distribution is of a double-step form resulting in smaller noise, characterized by $F=1/3$. Experimentally, however, the difference between cases (i) and (ii) is leveled by overheating of the non-ideal 2D leads~\cite{PhysRevB.59.2871}. Namely, monotonous growth of $T_L$ and $T_R$ with current makes the difference between local-equilibrium and double-step situations on the order of few percent in our experiment. Thus, in terms of input to the measured~$T_N$ in diffusive case situations (i) and (ii) are equivalent and we therefore limit ourselves to consideration of only the former case where we use eq.(\ref{localeq}).
\par
The noise of 2D leads in both ballistic and diffusive cases is calculated via eq.(\ref{BPF}). We treat 2D~leads and all formed $p-n$~junctions as uncorrelated noise sources and calculate the noise temperature of the whole sample using standard formalism~\cite{Blanter2001}. 

\subsubsection{Experimental results}
\begin{figure}[h]
\begin{center}
\includegraphics[width=0.9\columnwidth]{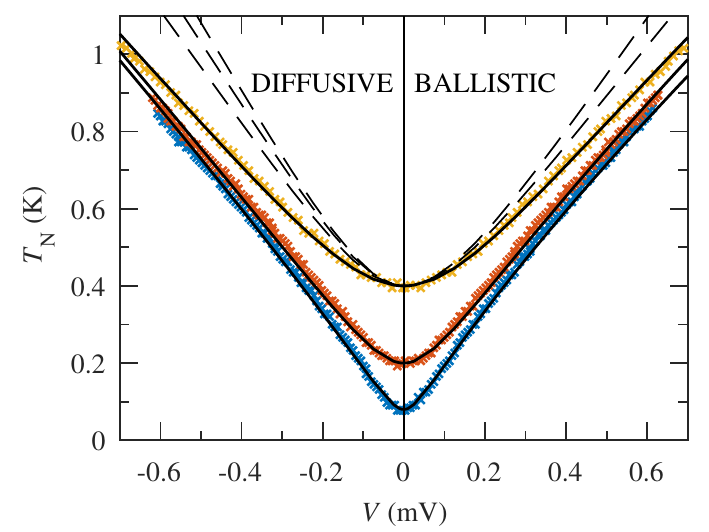}
\end{center}
\caption{Noise measurements for $p-n$~junctions at large negative gate voltage. Shot noise measured with contact N on the sample~S1 (symbols) at $T=80\,\mathrm{mK}$, $0.2\,\mathrm{K}$ and $0.4\,\mathrm{K}$ (from bottom to top) at $V_g=-6.5\,\mathrm{V}$. Solid lines are the best fits with $\Sigma_{\mathrm{h-ph}}(\mathrm{ball.})=0.1,\,0.08,\, 0.06\,\mathrm{W/m^2K^3}$ in the ballistic model (right half-panel) and $\Sigma_{\mathrm{h-ph}}(\mathrm{diff.})=2.5,\,2.5,\, 3\,\mathrm{W/m^2K^3}$ in the diffusive model (left half-panel), correspondingly. For the $T=0.4\,\mathrm{K}$ data by dashed lines we additionally plot model predictions for the values of~$\Sigma_{\mathrm{h-ph}}=0;\,0.014$ (equal to~$\Sigma_{\mathrm{e-ph}}$) and, in the case of diffusive model, for $0.06\,\mathrm{W/m^2K^3}$ which corresponds to the best fit in ballistic case.}\label{fig7}
\end{figure}

We now discuss experimental results obtained at large negative gate voltages. In fig.~\ref{fig7} we show the data (symbols) obtained on sample~S1 at~$V_g=-6.5\,\mathrm{V}$ at $T=80\,\mathrm{mK}$, $0.2\,\mathrm{K}$ and $0.4\,\mathrm{K}$ (from bottom to top; see Appendix~\ref{pnjn} for results obtained for sample S2). These results are almost $V_g$-independent for large negative $V_g$ (see also fig.~\ref{fig4}b). We emphasize that the data slope $d(2k_BT_N)/d(eV)$, lying in the range $0.15-0.25$ in this case, should not be confused with the Fano-factor of the $p-n$~junctions since the measured signal is comprised of both the junctions and the contacts noise inputs.
\par
As already discussed, in these measurements the $p-n$~junctions are connected to the ohmic contacts not only via 2D~leads with n-type conduction, but in part also via more resistive p-type conduction region under the gate. Unfortunately, the geometry of our samples doesn't allow to independently determine the rate of hole-phonon energy relaxation~$\Sigma_{\mathrm{h-ph}}$. We are not aware of any systematic studies of~$\Sigma_{\mathrm{h-ph}}$ in HgTe QWs, however, there are some indications from transport-based measurements of energy relaxation that the hole-phonon energy relaxation is significantly enhanced compared to~$\Sigma_{\mathrm{e-ph}}$~\cite{Sherstoprivate}. In the following, we compare experimental data with model predictions for various values of~$\Sigma_{\mathrm{h-ph}}$.
\par
The best fits of experimental data are shown by solid lines in fig.~\ref{fig7} for both ballistic (right half-panel) and diffusive (left half-panel) cases. Specifically, for the data at~$T=0.4\,\mathrm{K}$ we find that it is described by the ballistic model at~$\Sigma_{\mathrm{h-ph}}=0.06\,\mathrm{W/m^2K^3}$ and by the diffusive model at~$\Sigma_{\mathrm{h-ph}}=3\,\mathrm{W/m^2K^3}$. Additionally, for this data we plot model results for the values of~$\Sigma_{\mathrm{h-ph}}=0;\,0.014$ (equal to ~$\Sigma_{\mathrm{e-ph}}$) and, in the case of diffusive model, for $0.06\,\mathrm{W/m^2K^3}$, which corresponds to the best fit in ballistic case (dashed lines, from bottom to top). We find that for the values of~$\Sigma_{\mathrm{h-ph}}\approx \Sigma_{\mathrm{e-ph}}$ the model prediction is much greater than experimentally observed value even in the ballistic case, indicating that~$\Sigma_{\mathrm{h-ph}}\gg \Sigma_{\mathrm{e-ph}}$. This observation is general and holds for all temperatures and both samples. We summarize all obtained results in the table (all values are in the units of $\mathrm{W/m^2K^3}$).
\par
\begin{tabular}{ |c|c|c|c|c| }
\hline
\multicolumn{5}{ |c| }{Noise measurements} \\
\hline
Sample & T & $\Sigma_{\mathrm{e-ph}}$ & $\Sigma_{\mathrm{h-ph}}$ (ball) & $\Sigma_{\mathrm{h-ph}}$ (diff) \\ \hline
\multirow{6}{*}{S1} & 0.08 & 0.014 & 0.1 & 2.5 \\ 
 & 0.2 & 0.012 & 0.08 & 2.5 \\
 & 0.4 & 0.014 & 0.06 & 3 \\
 & 0.5 & 0.016 & 0.14 & 140 \\
 & 0.6 & 0.014 & 0.08 & 5 \\
 & 0.8 & 0.012 & 0.08 & $>140$ \\ \hline
\multirow{2}{*}{S1/2} & 0.1 & 0.016 & 0.14 & 14 \\
 & 0.2 & 0.016 & 0.14 & 14 \\ \hline
\multirow{3}{*}{S2} & 0.06 & 0.014 & 0.3 & 2.5 \\
 & 0.15 & 0.015 & 0.3 & 2.5 \\
 & 0.5 & 0.018 & 0.35 & 8 \\ 
\hline
\end{tabular} 
\par
As one can see, for the ballistic case the model is in agreement with experimental data for the relatively reasonable, almost $T$-independent, similar for all measured configurations values of~$\Sigma_{\mathrm{h-ph}}(\mathrm{S1})\lesssim0.14\,\mathrm{W/m^2K^3}$ and~$\Sigma_{\mathrm{h-ph}}(\mathrm{S2})\approx0.3\,\mathrm{W/m^2K^3}$. For the diffusive case these values must be additionally at least $10$~times greater, demonstrating also no consistency for various temperatures and samples. Although not providing definite value for the shot noise of $p-n$~junctions, these findings suggest that it is suppressed compared to the diffusive value.
\subsection{Energy relaxation of the edge states}
\begin{figure}[h]
\begin{center}
\includegraphics[width=0.9\columnwidth]{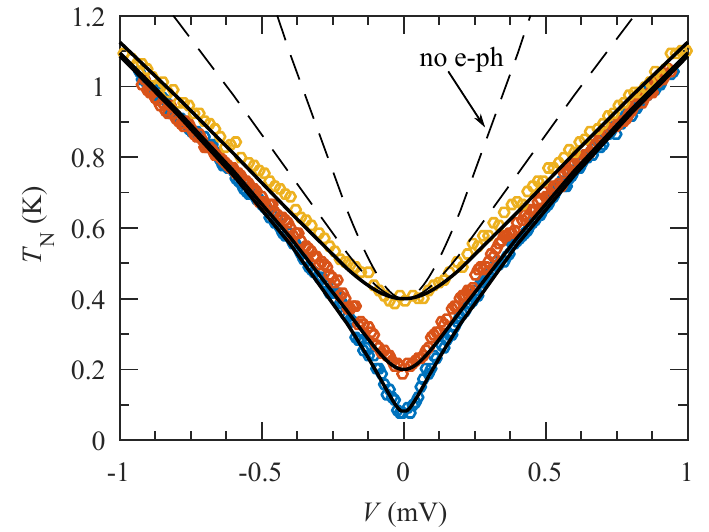}
\end{center}
\caption{Energy relaxation of the edge states. Noise measurements for several micrometers long edges N-C1 and N-C8, connected in parallel (see fig.~\ref{fig1}a), at bath temperatures $T=80\,\mathrm{mK}$, $0.2\,\mathrm{K}$ and $0.4\,\mathrm{K}$ on sample S1 at $V_g=-3.6\,\mathrm{V}$. Solid lines are plotted for $\Sigma_0=0.7\,\mathrm{\mu W/mK^3}$. For $T=0.4\,\mathrm{K}$ by dashed lines we additionally demonstrate the influence of edge energy relaxation for $\Sigma_0=0;\,0.3\,\mathrm{\mu W/mK^3}$.}\label{fig8}
\end{figure}
\par
While the $p-n$~junctions are relatively short, nevertheless in the diffusive case there is a possibility of hitherto neglected energy relaxation in the edge states themselves. It's rate~$\Sigma_0$ can in principle be inferred from the noise measurements on different length edges in a suitable geometry. We provide an estimate of~$\Sigma_0$ based on the measurements on several micrometers long edges in our samples assuming diffusive local-equilibrium situation. In fig.~\ref{fig8} we plot the noise temperature as a function of bias voltage for two -- $5\,\mathrm{\mu m}$ and $10\,\mathrm{\mu m}$-long edges connected in parallel, at bath temperatures $T=80\,\mathrm{mK}$, $0.2\,\mathrm{K}$ and $0.4\,\mathrm{K}$ (symbols, from bottom to top). Dashed upper line, plotted for $T=0.4\,\mathrm{K}$, corresponds to negligible inelastic scattering and demonstrates the importance of relaxation in the edge to the external bath. This is also suggested by sub-linear behavior of lowest-$T$ data $T_N(V)$ at low bias. In the given bias voltage range it is not possible to discern the exponent of energy relaxation of the edge states and we stick to $\alpha=3$ (which doesn't influence our general conclusion). We obtain an upper bound for the edge energy relaxation rate, expressed per unit length as $\Sigma_0=0.7\,\mathrm{\mu W/(mK^3)}$ (see solid lines in fig.~\ref{fig8}). To demonstrate the effect of $\Sigma_0$ by the lower dashed line we also plot the result of simulation for the intermediate value $\Sigma_0=0.3\,\mathrm{\mu W/mK^3}$.
\par
We now take energy relaxation in the $p-n$~junctions into account under the assumption of trivial diffusive conduction mechanism. With the value of $\Sigma_{\mathrm{h-ph}}$ corresponding to the best fit in the ballistic case, we find that $\Sigma_0=0.7\,\mathrm{\mu W/mK^3}$ in the diffusive case would be sufficient to fit experimental data of fig.~\ref{fig7} were the length of the $p-n$~junctions $L_{\mathrm{pn}}=1.6\,\mathrm{\mu m}$, which is much longer than electrostatically expected value $\sim~200\,\mathrm{nm}$. In Appendix~\ref{ER} we additionally demonstrate the influence of energy relaxation in the edge for various lengths of the $p-n$~junctions. Specifically, for $200\,\mathrm{nm}$-long $p-n$~junctions we find that the effect of edge energy relaxation with $\Sigma_0=0.7\,\mathrm{\mu W/mK^3}$ is almost negligible and leads to only slight correction $\Sigma_{\mathrm{h-ph}}=3\to2\,\mathrm{W/m^2K^3}$.
\section{Conclusion}
In summary, we studied the lateral $p-n$~junctions, electrostatically defined in 14\,nm--wide HgTe-based QWs. Resistance values close to~$h/2e^2$ and linear current-voltage characteristics indicate that transport across the junctions may be realized via two ballistic helical edge channels on either side of the junction. Noise measurements and numerical modelling are used to distinguish between the helical states and possible  trivial diffusive edge states scenarios. We take into account the effects of contacts heating, hole-phonon coupling in the p-type conduction region and energy relaxation in the edge states. Due to the unknown energy-relaxation rates our conclusion is not definitive, however, the obtained results are more consistent with the prediction for helical edge states transport across the $p-n$~junction. Our approach may allow
one to infer the edge transport mechanism in samples designed more suitably for noise measurements or provided the inelastic rates are known.
\begin{acknowledgments}
We thank D.V.\,Shovkun, A.A.\,Sherstobitov and A.I.\,Berdyugin for fruitful discussions. Experimental research at MIPT was supported by the RSF Project No.~15-12-30030. Experimental research at ISSP was supported by the RFBR Grant No.~15-02-04285 and numerical analysis was done under the support of RSF-DFG grant No.~16-42-01050. S.U.P. and E.S.T. acknowledge support from the RFBR Grant No.~16-32-00869.
\end{acknowledgments}
\appendix
\section{Noise in non-uniform case}
\label{BP}
\begin{figure}[h]
\begin{center}
\includegraphics[scale=0.7]{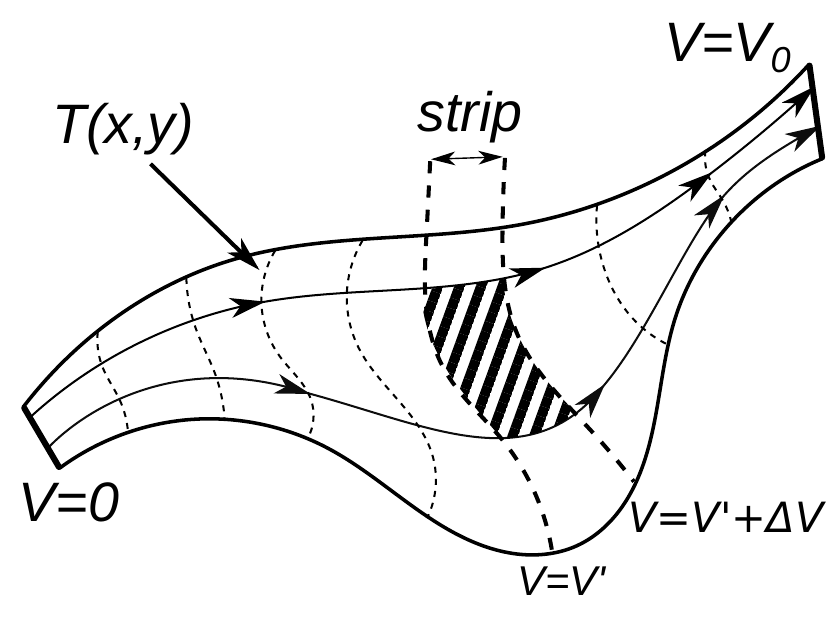}
\end{center}
\caption{Non-uniform 2D conductor.}\label{fig9}
\end{figure}
\par
We derive noise temperature~$T_N$ of a non-uniform 2D conductor (see fig.~\ref{fig9}), where fluctuations at different points are uncorrelated and can be represented by local noise temperature $T(x,y)$. The conductor is sliced into strips, separated by equipotential lines $V=\mathrm{const}$ (dashed lines on a figure). Such strips can be considered as resistors connected in series. Thus, the voltage fluctuations $S_V$ of the whole conductor is:
\begin{equation*}
S_V = \sum_{\mathrm{strips}} S_{V,\mathrm{strip}} = \sum_{\mathrm{strips}} S_{I,\mathrm{strip}} R_{\mathrm{strip}}^2.
\end{equation*}
Each strip is then divided into cells by current streamlines (lines with arrows on a figure) so, that no current is flowing through the boundaries between cells. Such cells can be considered as parallel resistors, so:
\begin{equation*}
S_{I,\mathrm{strip}} = \sum_{\mathrm{cells}} {S_{I,\mathrm{cell}}} = \sum_{\mathrm{cells}} { \frac{4k_B T_{\mathrm{cell}}}{R_{\mathrm{cell}}} }.
\end{equation*}
The noise temperature of the conductor $T_N$ is then determined by:
\begin{equation*}
T_N = \frac{S_V}{4k_B R} = \frac{1}{R} \sum_{\mathrm{strips}} R_{\mathrm{strip}}^2 \sum_{\mathrm{cells}} \frac{T_{\mathrm{cell}}}{R_{\mathrm{cell}}}.
\end{equation*}
The resistance of a cell can be expressed as $R_{\mathrm{cell}}=\Delta V/I_{\mathrm{cell}}$, where $\Delta V$ is a voltage drop on a strip and $I_{\mathrm{cell}}$ is a current flowing through the corresponding cell, and resistance of a strip can be expressed as $R_{\mathrm{strip}}=\Delta V/I$, where $I$ is a full current flowing through conductor. Then:
\begin{equation*}
T_N = \frac{1}{I^2 R} \sum_{\mathrm{strips}} \sum_{\mathrm{cells}} {T_{\mathrm{cell}} I_{\mathrm{cell}} \Delta V }.
\end{equation*}
Here $I_{\mathrm{cell}} \Delta V$ is a Joule power dissipated in a cell $Q_{\mathrm{cell}}$ and $I^2 R$ is a full Joule power dissipated in the conductor. In the limit we obtain
\begin{equation*}
T_N=\frac{\int T(x,y)\, Q_{\mathrm{Joule}}\,\mathrm{d}x\mathrm{d}y}{\int Q_{\mathrm{Joule}}\,\mathrm{d}x\mathrm{d}y},
\end{equation*}
where $Q_{\mathrm{Joule}}$ is the local Joule heat power.
\section{Noise of $p-n$ junctions}
\label{pnjn}
\begin{figure}[h]
\begin{center}
\includegraphics[width=0.9\columnwidth]{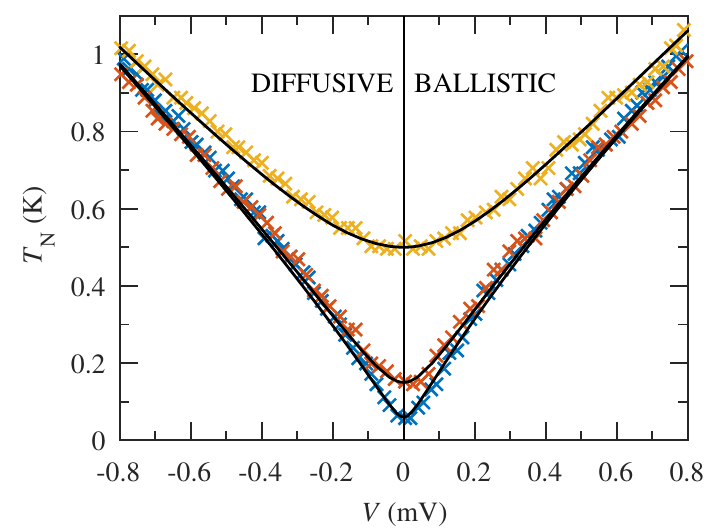}
\end{center}
\caption{Noise measurements for $p-n$~junctions at large negative gate voltage. Shot noise measured with contact N on the sample~S2 (symbols) at $T=60\,\mathrm{mK}$, $0.15\,\mathrm{K}$ and $0.5\,\mathrm{K}$ (from bottom to top) at $V_g=-7$; $-5.4$ and $-7\,\mathrm{V}$. Solid lines are the best fits with $\Sigma_{\mathrm{h-ph}}(\mathrm{ball.})=0.3,\,0.3,\, 0.35\,\mathrm{W/m^2K^3}$ in the ballistic model (right half-panel) and $\Sigma_{\mathrm{h-ph}}(\mathrm{diff.})=2.5,\,2.5,\, 8\,\mathrm{W/m^2K^3}$ in the diffusive model (left half-panel), correspondingly.}\label{fig10}
\end{figure}

In fig.~\ref{fig10} we show the data (symbols) obtained on sample~S2 at~$V_g=-7$; $-5.4$ and $-7\,\mathrm{V}$ at $T=60\,\mathrm{mK}$, $0.15\,\mathrm{K}$ and $0.5\,\mathrm{K}$, correspondingly (from bottom to top).
\section{Differential resistance}
\label{DR}
\begin{figure}[h]
\begin{center}
\includegraphics[width=0.9\columnwidth]{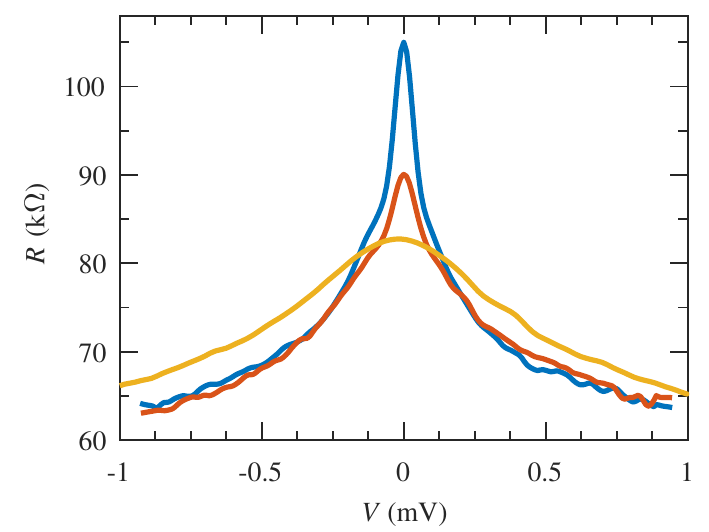}
\end{center}
\caption{Differential resistance of two -- $5\,\mathrm{\mu m}$ (N-C1) and $10\,\mathrm{\mu m}$-long (N-C8) edges connected in parallel on sample S1 (see fig.~\ref{fig1}a), at $V_g=-3.6\,\mathrm{V}$ at bath temperatures $T=80\,\mathrm{mK}$ (blue), $0.2\,\mathrm{K}$ (red) and $0.4\,\mathrm{K}$ (yellow).}\label{fig11}
\end{figure}

In fig.~\ref{fig11} we plot the differential resistance of two -- $5\,\mathrm{\mu m}$ (N-C1) and $10\,\mathrm{\mu m}$-long (N-C8) edges connected in parallel on sample S1. These data are obtained by numerical differentiation of I-V curves measured at $V_g=-3.6\,\mathrm{V}$. At sufficiently high bias the resistance at $T=0.4\,\mathrm{K}$ is slightly greater than at lower temperatures, probably due to some drift of $R(V_g)$-dependence on a large time scale.
\section{Energy relaxation of the edge states}
\label{ER}
\begin{figure}[h]
\begin{center}
\includegraphics[width=0.9\columnwidth]{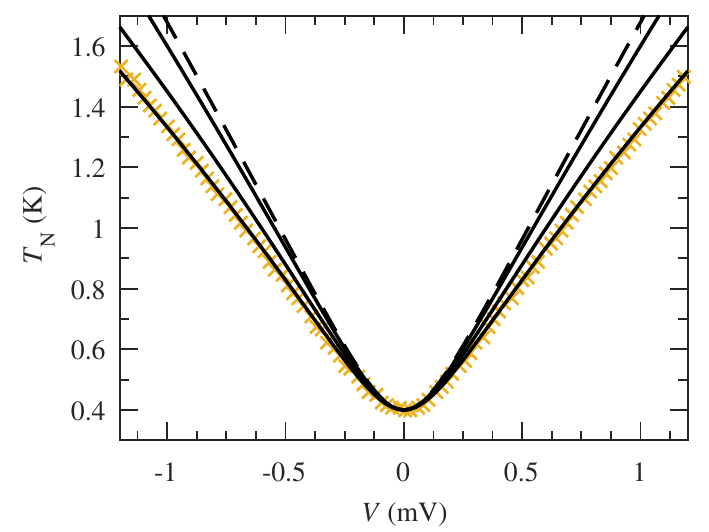}
\end{center}
\caption{Shot noise measured with contact~N on the sample~S1 (symbols) at $T=0.4\,\mathrm{K}$ at $V_g=-6.5\,\mathrm{V}$ (see fig.~\ref{fig7}). The dashed line reflects no energy relaxation in the $p-n$~junctions. Solid lines (top to bottom) are the diffusive model results for $L=0.2$, $0.8$ and $1.6\,\mathrm{\mu m}$ with $\Sigma_{\mathrm{h-ph}}=0.06\,\mathrm{W/m^2K^3}$, corresponding to the best fit of the ballistic model.}\label{fig12}
\end{figure}

Assuming trivial conduction diffusive mechanism with $\Sigma_{\mathrm{h-ph}}$ corresponding to the best fit in the ballistic case, we study if the distinction between ballistic and diffusive models may be ascribed to the energy relaxation in the $p-n$~junctions. In fig.~\ref{fig12} we demonstrate the influence of edge energy relaxation with the rate~$\Sigma_0=0.7\,\mathrm{\mu W/mK^3}$ (see fig.~\ref{fig8}) for various lengths of the $p-n$~junctions. We find that for the electrostatically expected value of the length of the $p-n$~junctions $L_{\mathrm{pn}}\approx200\,\mathrm{nm}$ the effect of energy relaxation is insignificant (see the top solid line). Our model shows that the experimental data would be consistent with the diffusive model only for the unreasonably longer $p-n$~junctions $L_{\mathrm{pn}}\approx1.6\,\mathrm{\mu m}$ (bottom solid line).

%

\end{document}